\documentclass[%
 reprint,
 amsmath,amssymb,
aps,
prx,
floatfix,
]{revtex4-2}

\usepackage{graphicx}
\usepackage{dcolumn}
\usepackage{bm,xcolor}
\usepackage{epstopdf}
\usepackage{hyperref}

\usepackage{amsmath,amssymb}

\newcommand{\Reff}[1]{Ref.~\onlinecite{#1}}

\def\eb{\begin{equation}}   
\def\ee{\end{equation}}     
\def\ea#1{\begin{eqnarray} #1 \end{eqnarray}}   

\def\mexp#1{\bigl< #1 \bigr>}

\def\ra{\rightarrow}

\def\of#1{\left(#1\right)}

\def\eq#1{Eq.~(\ref{#1})}

\newdimen\amstexex
\def\dotsadjustbox#1{\vbox to -1.4\amstexex{\kern-2\amstexex\hbox{#1}\vss}}

\begin{document}

\preprint{APS/123-QED}

\title{Efficient Evaluation of Exponential and Gaussian Functions on a Quantum Computer}

\author{Bill Poirier}
\email{Bill.Poirier@ttu.edu}%
\affiliation{Department of Chemistry and Biochemistry and Department of Physics and Astronomy,
Texas Tech University, Box 41061, Lubbock, Texas 79409-1061,
USA}%


\date{\today}

\begin{abstract}
The exponential and Gaussian functions are among the most fundamental and important operations, appearing ubiquitously 
throughout all areas of science, engineering, and mathematics. Whereas formally, it is well-known that any function may
 in principle be realized on a quantum computer, in practice present-day algorithms tend to be very expensive. 
In this work, we present algorithms for evaluating exponential and Gaussian functions efficiently on quantum computers.  
The implementations require a (generally) small number of multiplications, which represent the overall computational
bottleneck.  For a specific, realistic NISQ application,  the Toffoli count of the exponential function is found to be 
reduced from 15,690 down to 912,  when compared against a state-of-the art competing method by  H\"aner and coworkers 
[arXiv:1805.12445], under the most favorable  conditions for each method. For the corresponding Gaussian function comparison, 
the Toffoli count is reduced from 19,090 down to 704. Space requirements are also quite modest, to the extent that the
aforementioned NISQ application  can be implemented with as few as $\sim$70 logical qubits. More generally,  the methods 
presented here could also be equally well applied  in a  fault-tolerant context,  using error-corrected multiplications, etc.  
\end{abstract}

\pacs{Valid PACS appear here}
\maketitle


\section{\label{sec:Intro}Introduction}

The exponential and Gaussian functions are among the most fundamental and important operations, appearing ubiquitously 
throughout all areas of science, engineering, and mathematics. Whereas formally, it is well-known that any function,
$f(x)$, may in principle be realized on a quantum 
computer \cite{abrams97,zalka98,nielsen,florio04b,kais,preskill18,alexeev19,haner18,sanders20}, 
in practice present-day algorithms tend to be very expensive. 

Currently, one of the best strategies for evaluating general functions---as exemplified by Ref.~\cite{haner18}---divides 
the domain, $x$, into a collection of non-intersecting subdomains.
The function, $f(x)$, is then approximated using a separate $d$'th order polynomial for each subdomain,
evaluated through a sequence of $d$ multiplication-accumulation
(addition) operations.  The quantum advantage comes from the fact that these polynomial evaluations 
can be performed in parallel across all subdomains at once (using conditioned determination of 
the coefficients for each subdomain polynomial).

The above parallel quantum strategy is completely general, conceptually elegant, and effective.  It can also  be
optimized in various ways---e.g., for a given target numerical accuracy, and/or to favor gate complexity (i.e. number of
quantum gates or operations) over space complexity (i.e, number of qubits), or vice-versa.  However, in the words
of the Ref. \cite{haner18} authors themselves:
\begin{quotation} 
{\em While these methods allow to reduce the Toffoli [gate] and qubit counts significantly, the resulting circuits are still quite expensive, especially in terms of the number of gates that are required.}
\end{quotation}

Part of the reason for the ``significant expense'' is the cost of the requisite quantum 
multiplications \cite{draper00,florio04,florio04b,haner17,haner18,sanders20,parent17,gidney19,karatsuba62,kowada06}, each of which---at least for the 
most commonly used ``schoolbook'' algorithms \cite{haner17,haner18,sanders20,parent17,gidney19}---requires a sequence of $n$ 
controlled additions \cite{draper00,cuccaro04,takahashi,takahashi09,parent17}. Here, $n$ is the 
number of bits needed to represent the summands using fixed-point arithmetic (which is presumed throughout this work). 
Each controlled addition introduces $O(n)$ gate complexity---implying an overall quantum multiplication
gate complexity that scales as $O(n^2)$.
Although alternative multiplication algorithms with asymptotic scaling as low as $O(n^{\log_2 3})$ do 
exist \cite{parent17,gidney19,karatsuba62,kowada06},
they do not become competitive until $n$ reaches a few thousand. This is far beyond the values needed for most practical
applications (e.g., those of this work, for which $n=21$--32). 

For practical applications, then, there appear to be two  strategies that might be relied upon to 
significantly improve performance. The first is to wait for better quantum multiplication algorithms to be devised;
this is, after all, an area of active and ongoing development,  more so than general function evaluation on quantum
computers.  The second is to design entirely new algorithms, customized for \emph{specific} $f(x)$ functions.      

The present work is of the latter variety.  In particular, we present quantum algorithms designed  specifically to evaluate
exponential and Gaussian functions efficiently on quantum computers.  These algorithms require a (generally) small 
number of multiplications, which represent the overall computational bottleneck.  Our general approach is thus 
equally applicable to noisy intermediate-scale quantum (NISQ) calculations  \cite{preskill18,bharti21} 
with non-error-corrected quantum multiplications, as it is in a fault-tolerant context,  using error-corrected 
multiplications, etc. In all such contexts, we advocate for using the ``total multiplication count''  as the appropriate 
gate complexity metric---although the ``Toffoli \cite{nielsen} count'' metric, which is currently quite popular, 
 will also be used in this paper. 

It will be shown that the gate complexity for the present approach is dramatically reduced, when compared with the 
state-of-the art competing method by  H\"aner et al. \cite{haner18}.   For a specific, realistic  NISQ application,  the Toffoli 
count of the exponential function is reduced from 15,690 down to 912, under the most favorable 
conditions for each method. For the corresponding Gaussian function comparison, the Toffoli count is reduced
from 19,090 down to 704.  Space requirements are also generally reduced, and in any event quite modest---to 
the extent that  in one case, the above NISQ application  can be implemented with as few as 
$\sim$70 logical qubits. 

Although the range of applications where exponential and Gaussian functions are relevant is virtually limitless,
one particular application area will be singled out for further discussion. Quantum computational chemistry 
(QCC) \cite{poplavskii75,feynman82,lloyd96,abrams97,zalka98,lidar99,abrams99,nielsen,aspuru05,kassal08,whitfield11,brown10,christiansen12,georgescu14,kais,huh15,babbush15,kivlichan17,babbush17,babbush18,babbush18b,babbush19,low19,kivlichan19,izmaylov19,parrish19,altman19,cao19,alexeev19,bauer20,aspuru20}---i.e., quantum 
chemistry simulations \cite{szabo,jensen,helgaker,RN544} run on quantum 
computers---has long been regarded as one of the first important scientific applications where quantum supremacy 
will likely be realized \cite{aspuru05,georgescu14,altman19,aspuru20}.  
Particularly for ``first-quantized'' or coordinate-grid-based QCC \cite{abrams97,zalka98,lidar99,abrams99,nielsen,aspuru05,kassal08,whitfield11,huh15,babbush15,babbush17,kivlichan17,babbush19,alexeev19,bauer20,aspuru20}, it becomes necessary 
to evaluate functions over a (generally) uniformly-distributed set of discrete grid points \cite{aspuru05,babbush18,babbush19,aspuru20,RN553,colbert92,szalay96,light00,poirier02dvrlj,littlejohn02b}---exactly 
of the sort that emerges in fixed-point arithmetic, as used here. 

Of course, the most natural function to arise in the QCC context is the inverse-square-root
function, $f(x) = x^{-1/2}$, representing  Coulombic interactions \cite{szabo,jensen,helgaker,RN544}. 
Even for a ``general function evaluator'' code, this specific case poses some special challenges---associated, e.g., with the 
singularity at $x=0$---that result in substantially increased computational expense.  On the other hand, the
alternative  Cartesian-component separated (CCS) approach, as developed recently by the author and 
coworkers \cite{jerke15,jerke18,jerke19,mypccp,bittner},  replaces
the inverse square root with a small set of Gaussians. Using the new exponential/Gaussian 
evaluator of this work, then, the CCS approach would appear to become a highly competitive contender for 
first-quantized QCC.  

The remainder of this paper is organized as follows. Mathematical preliminaries are presented in
Sec.~\ref{subsec:prelim}, followed by an exposition of our basic quantum exponentiation 
algorithm in Sec.~\ref{subsec:basicquantum}, and its asymptotic scaling   in Sec.~\ref{subsec:cost}. 
These are the core results, especially for 
long-term quantum computing. Secs.~\ref{sec:refined} and~\ref{sec:detailed} then give a detailed explanation
of various algorithmic improvements leading to reduced gate and space complexity, that will be of particular
interest for NISQ computing.  In particular, quantum circuits for two specific NISQ implementations are 
presented in Sec.\ref{sec:detailed}---one designed to minimize gate complexity (Sec.~\ref{subsec:mult}), and the
other, space complexity (Sec.~\ref{subsec:alternative}). Using the specific ``gate saving'' and ``space saving'' 
implementations of Sec.~\ref{sec:detailed}, a detailed numerical comparison with Ref.~\cite{haner18} is provided 
in Sec.~\ref{sec:analysis}. Finally, concluding remarks are presented in Sec.~\ref{sec:conclusions}.

\section{\label{sec:basic} Basic Method}

\subsection {\label{subsec:prelim} Mathematical preliminaries}
Consider the exponential function, 
\eb
	f(x') = \exp \of{- \alpha x'}.  \label{fexprime}
\ee
We wish to evaluate the
function over the domain interval, $x'_{{\rm min}} \le x' < x'_{{\rm max}}$.  Note that 
 $x'$ and $\alpha$ are presumed to be \emph{real-valued}. If $\alpha$ were pure imaginary, then 
 \eq{fexprime} would be unitary---i.e., the most well-studied special case in quantum computing \cite{nielsen}.
 But this is not the case here. 
Without loss of generality, we may restrict consideration to $\alpha > 0$. The negative
$\alpha$ case corresponds to the above, but  with $x' \rightarrow -x'$, 
$x'_{{\rm min}} \rightarrow - x'_{{\rm max}}$, and $x'_{{\rm max}} \rightarrow - x'_{{\rm min}}$.

Both the domain and the range of \eq{fexprime} are represented discretely,
using a finite number of qubits. For generality, we allow the number of domain qubits, $d$,
to differ from the number of range qubits, $n$.  In the first-quantized QCC context, for instance, 
the $d \ll n$ case arises very naturally (where `$\ll$' represents perhaps a factor of 3 or 4). 
More specifically, something like 100 grid points are needed to accurately 
represent each domain degree of freedom---although the function values themselves require
a precision of say, 6--10 digits.  Throughout this paper, we mainly focus on the $d \ll n$ 
case---although the $d=n$ special case is obviously also important, and will also be considered. 

The $d$ qubits used to represent the domain correspond to $2^d$ distinct grid points, 
distributed uniformly across the $x'$ interval, with grid spacing 
$\Delta = 2^{-d} (x'_{{\rm max}} - x'_{{\rm min}})$.  Such representations are typical in quantum
arithmetic, and imply fixed-point rather than floating-point implementations \cite{haner18,sanders20}.
Indeed, since fixed-point arithmetic is closely related to integer arithmetic, we find it convenient
to transform $x'$ to the unitless domain variable,
\eb
	x = {(x'-x'_{{\rm min}})  \over \Delta },  \label{xeqn}
\ee
such that the $x$ grid points become \emph{integers}, $x = \{0, 1, \ldots, 2^d -1 \}$. 
In terms of $x$, the function then becomes
\ea{
	f(x) & = & \exp[-\alpha (x'_{{\rm min}} + \Delta x)] \nonumber \\
	       & = & C A^x, \quad \mbox{where}  \label{fCA}  \\
	C & = &  \exp(-\alpha x'_{{\rm min}}) \quad \mbox{and} \quad A = \exp(-\alpha \Delta). \label{CA}
} 

Next, we define the following binary decomposition of the $x$ integers, in terms of
the $d$ individual qubits, $x_i$, with $0 \le i < d$ and $x_i = \{0,1\}$:
\eb
	x = \sum_{i=0}^{d-1} x_i 2^i   \label{xbin}
\ee
Note that increasing $i$ corresponds to \emph{larger} powers of 2; thus, the binary
expansion of the integer $x$ would be $x_{d-1} \cdots x_1 x_0$.  Put another way, the lowest
index values correspond to the rightmost, or least significant, digits in the binary expansion. 
This convention shall be adopted throughout this work. 

Substituting \eq{xbin} into \eq{fCA}, we obtain
\ea{
	f(x) & = & C  \of{A^{2^0}}^{x_0}  \of{A^{2^1}}^{x_1}  \cdots   \of{A^{2^{d-1}}}^{x_{d-1}} \nonumber \\
	       & = & C \,A_0^{x_0} \, A_1^{x_1} \, \cdots \, A_{d-1}^{x_{d-1}}, \quad \mbox{where} \label{fCAi} \\
	       A_i & = & A^{2^i} \label{Ai}
}
In this manner, exponentiation is replaced with a sequence of $d$ multiplications. Note from \eq{Ai} 
that $A_0=A$. As additional notation, we find it convenient to  introduce the quantities $C_{0 \le i \le d}$, 
through  the recursion relation $C_{i+1} = C_i A_i^{x_i}$, with $C_0=C$.   Thus, $C_d = f(x)$, and 
the other $C_{i<d}$ quantities represent  partial products in \eq{fCAi}. 

\begin{figure*}[htb]
 \centering
 \includegraphics[height=9cm]{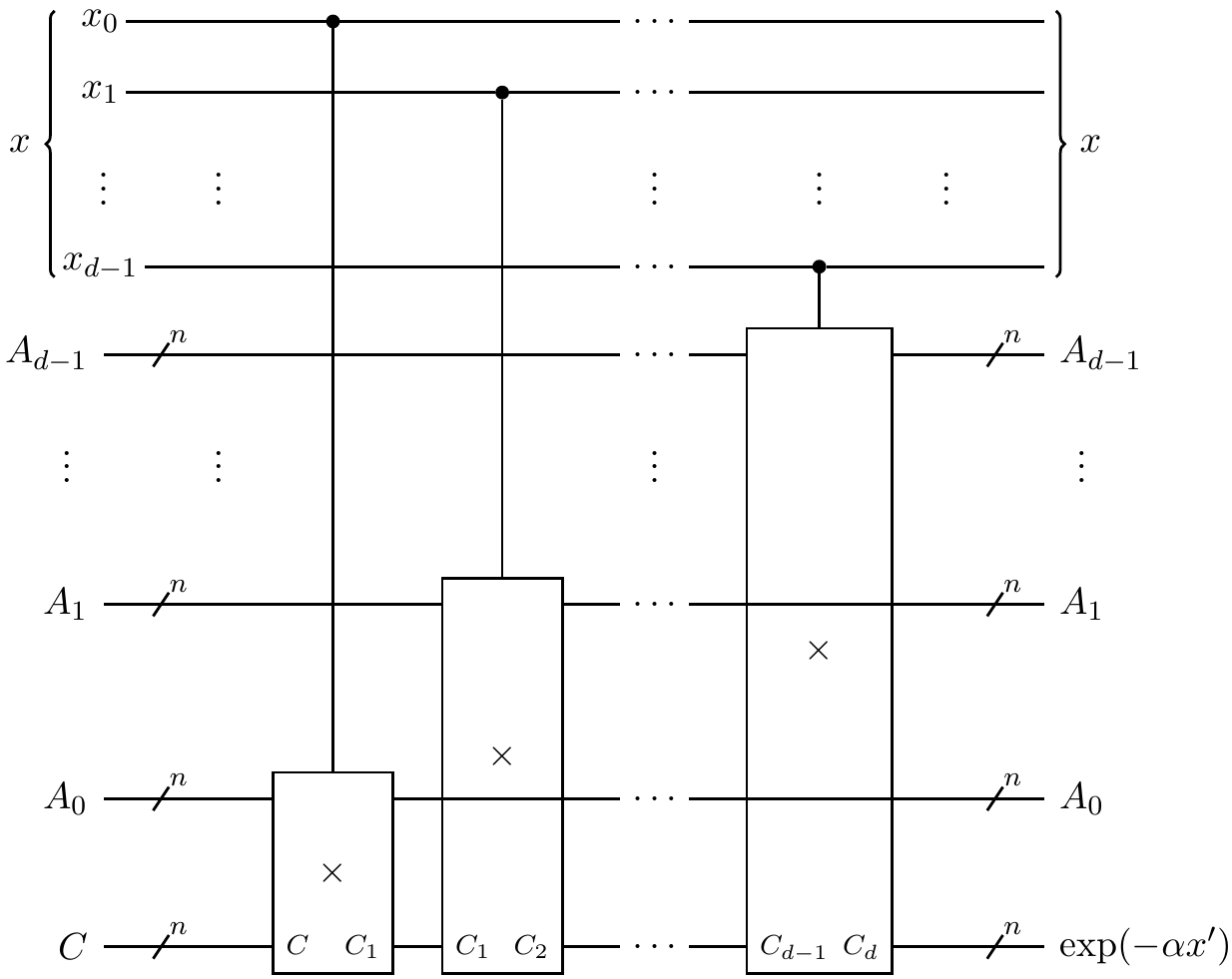}
 \vskip-5pt
 \caption{Quantum circuit used to implement basic quantum algorithm for exponentiation, 
 $f(x') = \exp(-\alpha x')$. All multiplications, $\times$, are presumed to be ``overwriting,''  in the sense
 that the second input register is overwritten with the product of the two inputs as output.}
 \label{fig1}
\end{figure*}

\subsection{\label{subsec:basicquantum} Basic quantum algorithm}

The exponent of every $A_i$ value in \eq{fCAi}, being the qubit $x_i$, is associated with the two 
states or values, 0 and 1.  From a quantum computing perspective, therefore, this situation
can be interpreted as an \emph{instruction}:
\begin{itemize}
\item If $x_i = 1$, then multiply by $A_i$.
\item Otherwise, i.e. if $x_i=0$, do nothing.
\end{itemize} 
This suggests a simple and straightforward quantum algorithm for exponentiation, 
consisting of nothing but a sequence of $d$ \emph{controlled multiplications}, as
 indicated in Fig.~\ref{fig1}. 

From the figure, each of the $d$ qubits, $x_i$, serves as the control qubit for a separate 
target multiplication of $C_i$ by $A_i$, in order to generate the next $C_{i+1}$.  
In this basic implementation, each of the $d$ constants, $A_i$, is stored by a separate
bundle of $n$ qubits, initialized prior to the calculation.  An additional bundle of $n$ 
qubits (lowest wire in Fig.\ref{fig1}) is used to represent the value of the function.  This output 
register is initially assigned the constant value $C$,  but through  successive controlled multiplications with $A_i$ 
as described above, ends up taking on the final output value,  $C_d=f(x) = \exp(-\alpha x')$. 

For the moment, we primarily treat multiplication as an oracle or ``black box'' routine,
whose operational details need not concern us. 
However, we note from the above description (and from  Fig.\ref{fig1}) that one of the 
two input registers gets overwritten with the product value as output, and the other is unaffected. 
There are indeed some multiplication algorithms---e.g. those based on the Quantum Fourier
Transform (QFT) \cite{shor94,abrams99,nielsen,draper00,florio04,florio04b}---that behave in this manner.  
We call these ``overwriting'' 
multiplication routines.  Other standard multiplication algorithms---e.g., those based on
bit-shifted controlled additions \cite{haner17,haner18,sanders20}---do not have this property. 
This issue is revisited again in Sec.~\ref{sec:detailed}. 

As discussed, the  $C_i$ values are stored in the $n$-qubit output register, whereas the
$A_i$ are stored in $d$ separate $n$-qubit  input registers. Since  
$0 < A_i < 1$ for all $A_i$, it is convenient to 
represent these constants using the following $n$-bit binary expansion:
\eb
	y = \sum_{j=0}^{n-1} y_j 2^{-(n-j)}   \label{ybin}
\ee 
Thus, the binary expansion of $y$ becomes $y=0.y_{n-1} \cdots y_1 y_0$---with the
$y_0$ bit least significant, as discussed.  This representation has a resolution of
$2^{-n}$.  Likewise, $0 < C_i  \le 1$ for all $0 \le i \le d$, provided 
$x'_{{\rm min}} \ge 0$ (if not, there are simple remedies that can  be applied, although
these are not needed here).  We therefore  find it convenient
to adopt the \eq{ybin} representation for the $C_i$  as well as the $A_i$ values. 

The above describes the basic algorithm for evaluating the exponential function
of \eq{fexprime}.  For the Gaussian function, i.e. 
\eb
	f(x') = \exp(-\alpha x'^2), \label{fexpGauss}
\ee
one proceeds in exactly the same manner, except that it  is necessary to perform an additional
multiplication, to obtain $x^2$ from $x$.  We note that there are some specialized 
quantum squaring algorithms, that shave a bit off of  the cost
of a generic multiplication \cite{haner18,sanders20,gidney19}.  If $d \ll n$ however, this  savings is not significant;
 the cost of the extra multiplication itself is much less than the others, since it involves
only $d$ rather than $n$ qubits.

\begin{figure*}[htb]
 \centering
 \includegraphics[height=6.5cm]{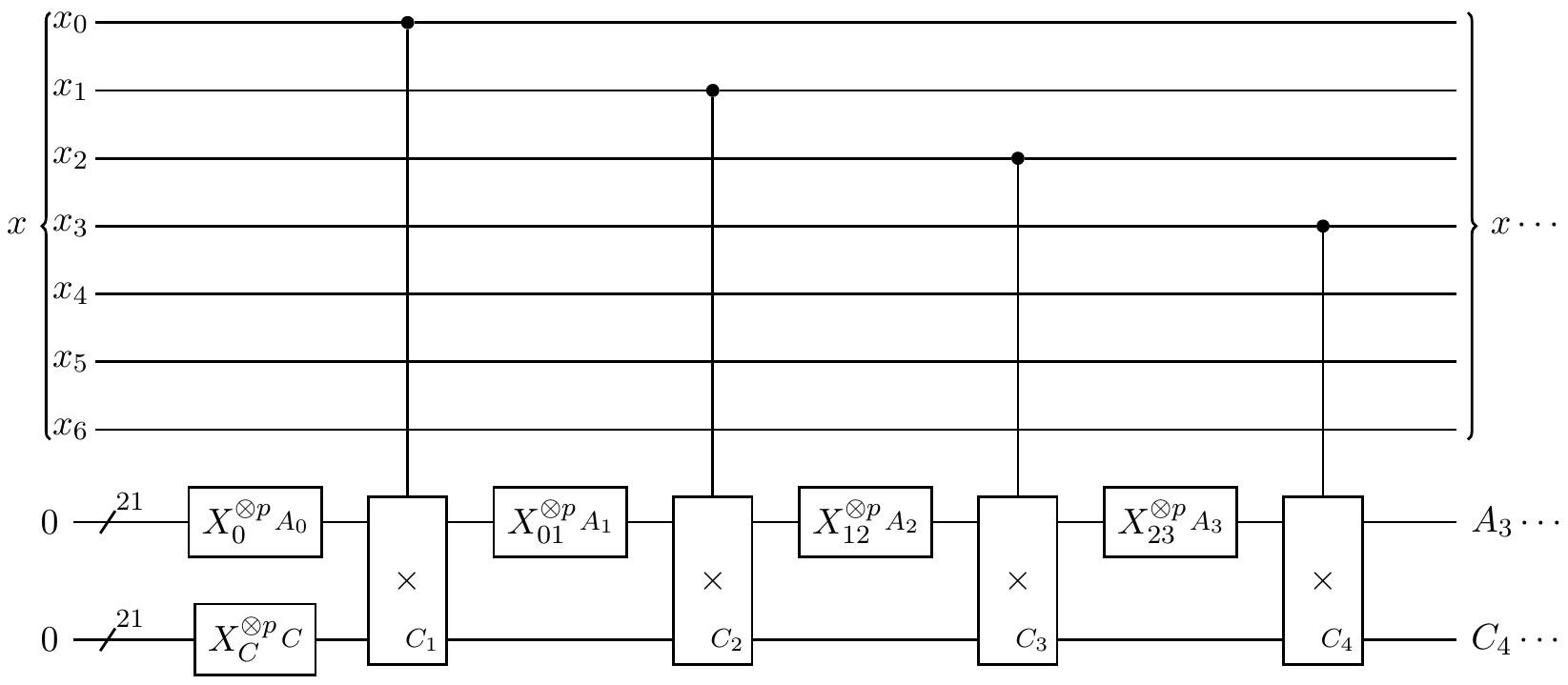}
 \vskip-5pt
 \caption{First half of quantum circuit used to implement refined quantum algorithm for exponentiation, 
 $f(x') = \exp(-\alpha x')$, for specific parameter values, $d=7$, $n=21$, and $A=0.389$.
 Overwriting multiplications are presumed.}
 \label{fig2}
\end{figure*}

\subsection{\label{subsec:cost} Computational cost and asymptotic scaling}

In terms of memory (i.e., space) usage, the above algorithm requires 
$dn + n + d$ qubits in all---not including the ancilla bits needed to actually
implement the multiplications (not shown in  Fig.~\ref{fig1}).   As mentioned, the 
computational cost is simply that of applying $d$ multiplications. Given the tremendous
variety of multiplication algorithms that have been and will be developed---and given
that some will always be better than others in different circumstances---\emph{we feel
it is best to let the number of required multiplications itself serve as the appropriate 
gate complexity metric}. Of course, this requires that multiplications comprise the 
overall computational bottleneck, as they do here.  In similar fashion, the Toffoli 
count  provides another implementation-independent metric---when 
comparing circuits whose bottleneck is the Toffoli gate (Sec.~\ref{sec:analysis}).

If absolute  costs are difficult to compare directly between different methods, then
the next best thing to consider is asymptotic scaling---in this case, in terms of the 
parameters $n$ and $d$.  For our basic exponentiation algorithm, the scaling with
respect to $d$ is clearly linear---both of the space 
and gate complexity.  

As for the scaling with respect to  $n$, this is determined by the multiplication algorithm
itself. At present, the most competitive quantum multiplication algorithm for asymptotically large $n$
in terms of scaling appears to be that of C. Gidney \cite{gidney19}, based on the 
recursive Karatsuba scheme \cite{karatsuba62}. The Gidney algorithm requires $O(n)$ space 
complexity, so that the overall scaling for our basic exponentiation algorithm  
would be $O(nd)$.  Likewise, the gate complexity for a single multiplication scales 
as $O(n^{\log_2 3})$, implying  $O(n^{\log_2 3} d)$ scaling for basic exponentiation.  

As mentioned, Gidney does not overtake even 
the simplest (i.e. ``schoolbook'') multiplication method until $n$ reaches a few thousand. 
It is therefore not practical for NISQ computing.  In 
Sec.~\ref{sec:analysis}, more precise estimates will be provided for absolute
costs---e.g. in terms of Toffoli counts---presuming multiplication methods that can
be  practically applied in a NISQ context (Sec.~\ref{sec:detailed}). We also improve upon the basic 
exponentiation algorithm itself---in Sec.~\ref{sec:refined}, where we adopt
a more efficient and refined approach, and in Sec.~\ref{sec:detailed}, where we present
a specific, NISQ implementation. 

At this point it is worthwhile to compare the two cases, $d=n$ and $d \ll n$. If the 
exponentiation operation is itself part of a  more complicated mathematical function network,
with many nested inputs and outputs, then presumably one wants a generic $d=n$ code with $n$
sufficiently large to provide ``machine precision''---i.e., $n\ge 25$ or so for single precision,
or $n \ge 50$ for double precision.  The $O(n^2)$ space complexity of our basic exponentiation
algorithm likely  places such calculations beyond the current NISQ frontier.  

On the other hand, there are situations where $d \ll n$, and where $n$ itself may be
substantially reduced.  For first quantized QCC, for example, it is estimated that $d=7$ 
and $n=21$ may suffice to achieve the so-called ``chemical accuracy'' benchmark \cite{mypccp}.
Such values place the present scheme much closer to the NISQ regime---especially
once the refinements of the next section are introduced. 

We conclude this subsection with a reexamination of the true cost of the   $d \ll n$ 
Gaussian function evaluation, within the present basic scheme. Though as stated,
the $x^2$ operation \emph{per se} adds little to the direct cost,  it does have the 
effect of squaring the size of the domain interval. Thus, if the full resolution 
of the domain is to be preserved, this requires $2d$ rather than $d$ qubits---as well
as a commensurate doubling of the gate complexity.   On the other hand, 
this relative increase can often be largely mitigated by the  improvements introduced in the 
subsequent sections. 

\section{\label{sec:refined} Refined Method}

The basic algorithm can be substantially improved, with respect to  both space and gate complexity,
using the refinements described in this section. For definiteness, going forward
we shall generally presume the ``NISQ parameter values,'' $d=7$ and $n=21$, as discussed in
Sec.~\ref{subsec:cost}.  However, for comparison and robustness testing, we shall occasionally
use the less spartan parameter values, $d=8$ and $n=32$ (corresponding to ``machine precision''). 
In both cases, we find that a NISQ calculation is likely feasible in the near-term future. 

There are essentially two distinct ideas presented  in this section to improve upon the basic 
algorithm---although other possible options certainly also exist.   The first idea is to \emph{transform}
the $A_i$ values between successive multiplications, so that only one such constant need be stored
at a time. This will have the effect of reducing the space complexity scaling to $O(n)$, at least for
overwriting multiplications.  The second idea reduces the actual number of multiplications that need be 
applied. 

\subsection{\label{subsec:ref1} Refinement \# 1: reducing space complexity}

The parameters $A_i $ have constant values that can be determined prior to the calculation.
Rather than storing them in $d$ separate registers, it is far less costly in terms of space 
to simply transform $A_i \rightarrow A_{i+1}$, prior to each successive multiplication. Such strategies
have been used previously in quantum computing, when constant (unsuperposed) register values 
are employed \cite{parent17,haner18,shor94}. If overwriting  multiplications are used, it then 
becomes necessary to maintain only two such $n$-qubit registers---i.e., one to store all of the  
successive $A_i$ values, and the other to store the (conditionally superposed) 
$C_i$ values. 

The corresponding quantum circuit is presented in Fig.~\ref{fig2}, for $d=7$ and 
$n=21$.  The upper of the two 21-qubit registers is used to store the $A_i$, with the transformation
gate $X_{i(i+1)}^{\otimes p}$ used to transform $A_i$ into $A_{i+1}$.  Similarly, we  define 
transformation gates $X_{i}^{\otimes p}$ to convert the zero state $0$ into $A_i$ (or vice-versa).  
For example, the gate $X_{0}^{\otimes p}$ is used at the start of the circuit to initialize $A_0$ from 0.
Likewise, the lower 21-qubit register is initialized to $C$ from 0, using the transformation gate
$X_{C}^{\otimes p}$.  Each successive multiplication operation (conditionally) multiplies this value by 
another factor of $A_i$.  
In this manner, the total number of qubits is reduced to $2n + d $, or 49 for the present NISQ 
example---again, not including the various ancilla bits needed to effect the (overwriting) multiplications
in practice.

The  strategy above emphasizes minimal space complexity at the cost of
greater gate depth.  Alternatively, using all $d$ $A_i$ registers as in Sec.~\ref{sec:basic}, 
the multiplications could be performed synchronously and hierarchically, so as to minimize
gate depth, but without any space reduction.
In any event, our analysis in Sec.~\ref{sec:analysis} is all based on 
\emph{non}-overwriting multiplications (Sec.~\ref{sec:detailed}), for which the situation is a bit more
complicated.

We next turn our attention to the implementation of the transformation gates. 
Since  the transformations always correspond to fixed input 
and output values,  they can easily be implemented as a set of very specific  \textsc{NOT} gates,
applied to just those qubits for which the binary expansions of \eq{ybin} differ between
input and output values. 
Hence the notation, `$X^{\otimes p}$', to refer to the resultant tensor product of $p \approx n/2$
 \textsc{NOT} gates used to effect the transformation.
 
\begin{figure}[h]
\centering
  \includegraphics[height=14cm]{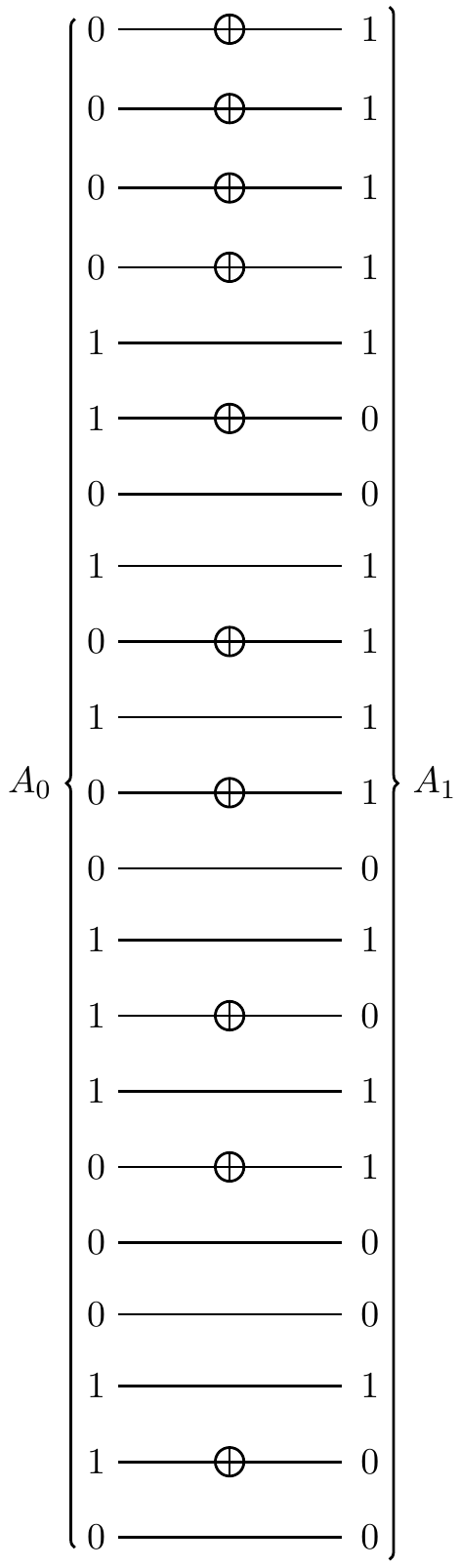}
  \vskip-5pt
  \caption{Quantum circuit used to implement $X_{01}^{\otimes p}$
on a quantum computer, for specific parameter values, $d=7$, $n=21$, and $A_0=A=0.389$. 
The $n=21$ binary representation of $A_0$ is $A_0=.011000111001010110000$;
that of $A_1=A^2 = 0.151321$ is $A_1=0.001001101011110011111$. The least significant bit, 
i.e. $j=0$, appears at the top of the circuit.  For this example, $p=10 \approx n/2$.}
  \label{fig3}
\end{figure}

 In Fig.~\ref{fig3}, the specific implementation for $X_{01}^{\otimes p}$ is presented, 
 corresponding to the specific values, $d=7$, $n=21$, and $A=0.389$.  The input qubits 
are in an unsuperposed state corresponding to the $n=21$ binary expansion of $A_0=A$,
as expressed in the form of \eq{ybin} (with $y_0$ corresponding to the top wire, etc.)  The 
output qubits are in a similar state, but corresponding to $A_1 = A^2 = 0.151321$. Generally
speaking, we may expect about half of the qubits to change their values. Indeed, for the
present example with $n=21$, we find $p=10$.  
  
In the refined algorithm as presented in Fig.~\ref{fig2}, we find that there is one transformation
required per multiplication.  However, it is clear from Fig.~\ref{fig3} that the gate complexity
of the former is trivial in comparison with that of the latter.  In practical terms, therefore, the scheme
of Fig.~\ref{fig2} can be implemented at almost no additional cost beyond that of  Fig.~\ref{fig1}---i.e., 
we can continue to use  multiplication count as the measure of gate complexity.

\begin{figure*}[htb]
 \centering
 \includegraphics[height=9.5cm]{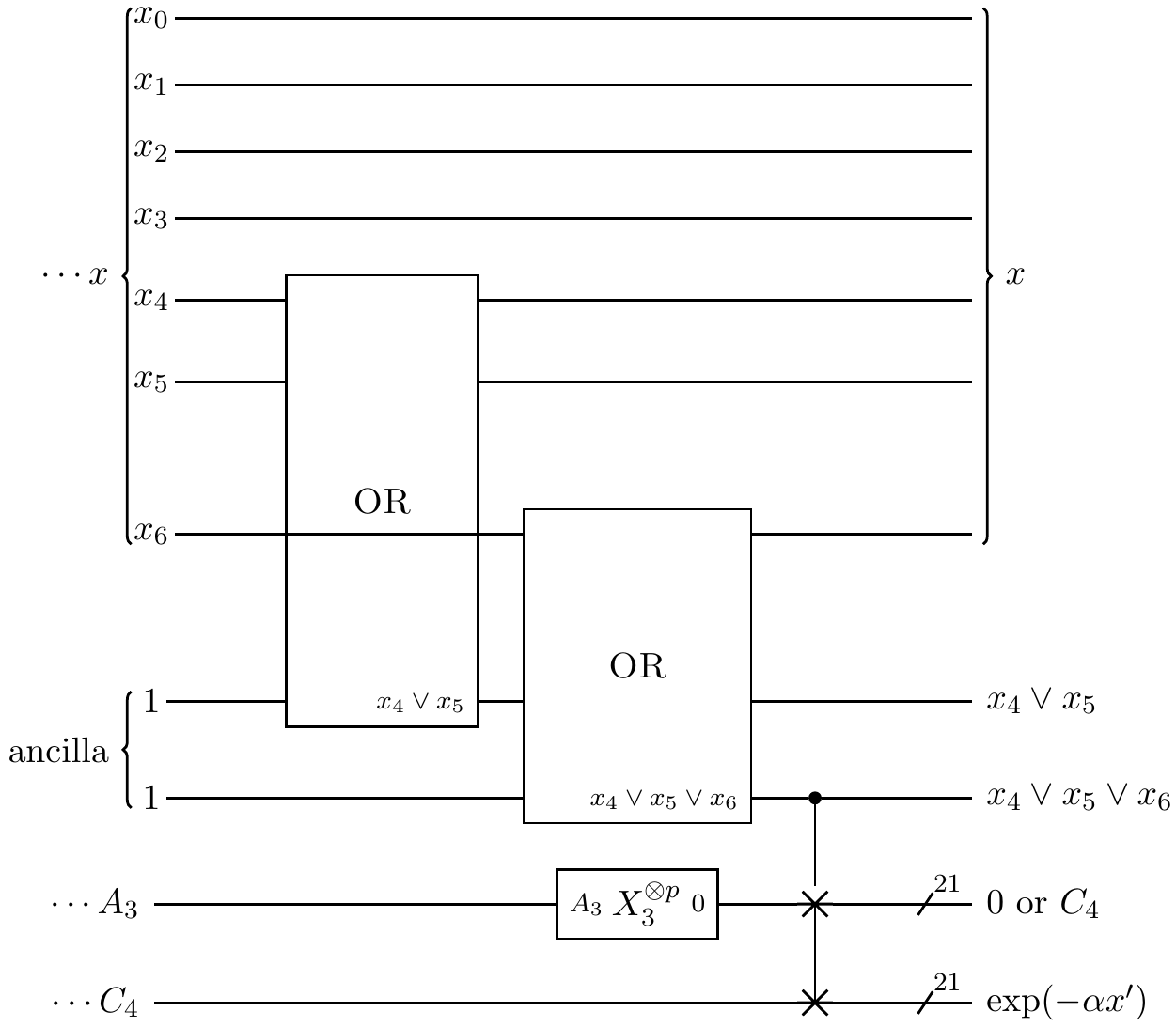}
 \vskip-5pt
 \caption{Second half of quantum circuit used to implement refined quantum algorithm for exponentiation, 
 $f(x') = \exp(-\alpha x')$, for specific parameter values, $d=7$, $n=21$, and $A=0.389$.}
 \label{fig4}
\end{figure*}

\subsection{\label{subsec:ref2} Refinement \# 2: reducing gate complexity}

In the initial discussion that follows, it is convenient to reconsider the $d=n$ case. 
Note that for both the basic quantum algorithm of Sec.~\ref{sec:basic}, and the refined
version of Sec.~\ref{subsec:ref1}, a total of $n$ multiplications are required---implying
an overall gate complexity that scales asymptotically as $O(n^{1+\log_2 3}) \approx O(n^{2.585})$,
if Karatsuba multiplication is used.
In reality, however, not all $n$ of the multiplications need be applied in practice. In
fact, it will be shown in this subsection that the actual required number of multiplications,
$m$, scales as $\log n$ (for fixed $A$)---thereby implying an asymptotic scaling of gate  complexity
no worse than $O(n^{\log_2 3} \log n)$.

The important realization here is that \eq{Ai} implies a \emph{very rapid} reduction in
$A_i$ with increasing $i$---essentially, as the exponential of an exponential.  
Consequently, there is no need to apply an explicit multiplication for any $A_i$ 
whose value is smaller than the smallest value that can be represented numerically
in our fixed-point representation---i.e.,
$2^{-n}$, according to \eq{ybin}.  What is needed, therefore, is an expression for $m$
in terms of $A$ and $n$, where $m$ is the smallest $i$ such that $A_i < 2^{-n}$. 

For the $d=n$ case, it can easily be shown that
\eb
	m = \left \lfloor  \log_2 \! \of{{n \over \log_2 (1/A)}}  \right \rfloor   + 1 . \label{meq}
\ee
For the generic case where $d$ and $n$ are independent, we still never need more
than $d$ multiplications. So  \eq{meq} above gets replaced with the general form,
\eb
	m = \min \left \{d,  \left \lfloor  \log_2 \! \of{{n \over \log_2 (1/A)}}  \right \rfloor   + 1 \right \}. \label{meq2}
\ee

Clearly, $m$ scales asymptotically as either $O(d)$ or $O(\log n)$, rather than $O(n)$, if $A$ is fixed.  
This assumes, however, that $d$ and $A$ have no implicit dependence on $n$, which in turn depends 
on assumptions about how the $x'$ grid points are increased.  If the $x'_{\rm min} \le x' < x_{\rm max}$ domain 
interval is expanded keeping the same spacing $\Delta$, \emph{or} if $\Delta$ decreases but $d$ is kept
constant, then the above holds.  Otherwise, $A \rightarrow 1$ as $n \rightarrow \infty$,  and the 
prefactor becomes divergently large, implying a less favorable asymptotic scaling law. 

Let us consider the case where $d < n$.    Since $m(A)$  as described by \eq{meq} increases 
monotonically with $A$,  there is in general an interval $0 < A < A_{{\rm max}}$ over which 
$m(A) < d$, and so a reduction in the number of multiplications can be realized and  $m<d$. Beyond 
this point---i.e., for $A_{{\rm max}} \le  A < 1$, all $m=d$ basic multiplications must be  used. 
A bit of algebra reveals the following expression for the transition $A$ value:
\eb
	A_{{\rm max}} = 2^{- n/2^{d-1} } \label{Amax}
\ee

As an illustrative example, consider the $d=7$, $n=21$, $A=0.389$ case of 
Sec.~\ref{subsec:ref1}.  The formula of \eq{meq} predicts that $m=4$ multiplications
will be required, exactly as indicated in Fig.~\ref{fig2}.  This represents a significant
reduction versus the $d=7$ multiplications that would otherwise be needed. As
confirmation that $m=4$ is correct, we note that $A_3=0.00052432$, which is
larger than $2^{-21} = 4.768 10^{-7}$.  However, $A_4 = 2.749 10^{-7} < 2^{-21}$. 

Finally, we can compute $A_{{\rm max}}$ from \eq{Amax}---which, with the above $n$ and $d$
values, is found to be $A_{{\rm max}} = 0.796571$.  Thus, one finds  a reduction in $m$
down from $d$, over about 80\% of the range of possible $A$ values. Now consider the
\emph{Gaussian} rather than exponential function, for which $d \ra 2d=14$. Here,
we find $A_{{\rm max}} = 0.998225$---implying that there is \emph{almost always} 
a reduction in $m$.  We will discuss further ramifications in Secs.~\ref{sec:detailed} and 
~\ref{sec:analysis}.

\subsection{\label{subsec:second} Second half of refined quantum algorithm}

Although the second refinement of Sec.~\ref{subsec:ref2}, can lead to fewer than $d$ 
multiplications (depending on the values of $n$, $d$, and $A$), this does not  
imply that the refined quantum algorithm simply ends at the right
edge of Fig.~\ref{fig2}.  There remains a subsequent computation that must occur, using   
the  $x_{i\ge m}$ qubits, in order to ensure that the correct final value 
for the function is obtained.  The  multiplication count of the additional computation
is zero, although it does add a cost of $n$ Fredkin gates.  

Consider that when $x=2^i$ is a power of two, then all but the $x_i$ binary expansion
coefficients in \eq{xbin} vanish, and the function value becomes simply $f(x) = C A_i$,
according to \eq{fCAi}. This implies that for any $x \ge 2^m$, $f(x) < 2^{-n}$ is smaller than the 
minimum non-zero number that can be represented, and so should be replaced with $f(x)=0$.  
This situation will occur if \emph{any} of the $(d-m)$ qubits, $x_{i\ge m}$, are in their 1 states. 
Otherwise---i.e., if all $(d-m)$ of the $x_{i\ge m}$ are in their 0 states so that $x < 2^m$---then 
nothing should happen, as the lowest register is already set to the correct output value, $f(x)=C_m$,
at the right edge of  Fig.~\ref{fig2}. 

The above can be implemented as follows. First, for the case  $A \ge A_{{\rm max}}$, 
no additional circuitry is needed; one simply runs the quantum circuit of Fig.~\ref{fig2} as is, except 
with explicit controlled multiplications across all $m=d$ of the $x_i$ qubits.  For the case $m=d-1$,
then $d-1$ controlled mutiplications are implemented across the lowest $d-1$ qubits,
$x_{i < (d-1)}$.  The final qubit, $x_{d-1}$ is then used to conditionally set the lowest 
register to zero. 

For the last case where $(d-m) \ge 2$, we apply the quantum circuit indicated in Fig.~\ref{fig4}.   
This requires first checking if any of the  $(d-m)$ $x_{i\ge m}$ qubits are in state 1, 
which is implemented using a sequence of $(d-m-1)$ \textsc{OR} gates.  The first
is applied to $x_m$ and $x_{m+1}$ to compute $x_m \vee x_{m+1}$. If needed, that 
output is then sent to a second \textsc{OR} gate along with $x_{m+2}$, etc.  The final
output, which will serve as a control qubit, has value 1 if any of the $x_{i\ge m}$
are in their 1 states; otherwise, it has value 0. 

Meanwhile, the upper of the two $n$-qubit registers, which starts out representing
the constant value $A_{m-1}$, is transformed to the value 0, using the transformation
gate, $X_{m-1}^{\otimes p}$.  Finally, the upper and lower $n$-qubit registers undergo
a controlled $\textsc{SWAP}^{\otimes n}$, applied in qubit-wise or tensor-product fashion, 
across all $n$ qubits of the two registers.  If the swap occurs, then the function output
as represented by the lower of the two $n$-qubit registers becomes zero; 
otherwise, it is left alone. 

\begin{figure}[h]
\centering
  \includegraphics[height=1.8cm]{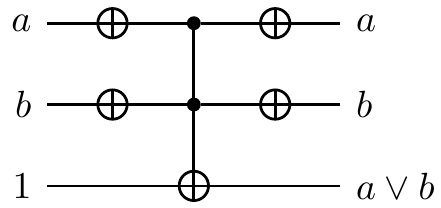}
  \vskip-5pt
  \caption{Quantum circuit used to implement reversible \textsc{OR} gate, constructed out of 
  a single reversible \textsc{NAND}  (Toffoli) gate, and various \textsc{NOT} gates.}
  \label{fig5}
\end{figure}

We conclude this subsection with a discussion of the reversible quantum \textsc{OR} gate, 
used in the quantum circuit of Fig.~\ref{fig4}. Such
a gate can be easily constructed from a single reversible \textsc{NAND} (Toffoli) gate, 
together with various \textsc{NOT} gates, as indicated in Fig.~\ref{fig5}. Note 
that each such \textsc{OR} gate introduces one new  ancilla qubit, initialized to the 1 state. 
There are thus no more than  $(d-m-1)$ additional ancilla qubits in all that get introduced in 
this fashion.  The additional costs associated with Fig.~\ref{fig4}, in terms of both gates
and qubits, are thus both very small as compared to those of Fig.~\ref{fig2}, although 
they will be included in resource calculations going forward.

\section{\label{sec:detailed} Detailed Implementation Suitable for NISQ Computing}

\subsection{\label{subsec:overview}  Overview}

As discussed, there is  large variety of quantum multiplication algorithms
on the  market currently  
\cite{draper00,florio04,florio04b,haner17,haner18,sanders20,parent17,gidney19,karatsuba62,kowada06}, 
and no doubt many more will follow. 
In part for this reason, we prefer to rely on the ``multiplication count'' metric for gate complexity. Indeed,
 whereas current multiplication algorithms largely make use of integer or  fixed-point 
arithmetic, floating-point algorithms---which have very different implementations---are also 
of  interest going forward, especially for exponentiation.   The multiplication 
count metric will continue to be relevant for all such innovations. 

On the other hand, we are also interested in developing a specific exponentiation circuit
that can be run on NISQ computers for realistic applications. Moreover, we aim
to compare performance against the state-of-the-art competing method by H\"aner and 
coworkers \cite{haner18}, for which multiplications are not the only bottleneck. 
 This constrains us in two important ways.
First, we cannot use the multiplication count metric for accurate comparison; instead, since the  
H\"aner algorithm is Toffoli-based (as is our circuit),  we use the Toffoli count metric. 
Second, to the extent that both exponentiation algorithms do
rely on multiplications, similar multiplication subroutines should be used for both. 

Accordingly, we use a modified version of H\"aner's multiplication subroutine, which is itself a
fixed-point version of  ``schoolbook'' integer multiplication \cite{haner17,haner18,sanders20,parent17,gidney19}, 
based on bit shifts and controlled additions. In particular, they exploit truncated additions (that 
maintain $n$ fixed bits of precision), together with a highly efficient overwriting, controlled, 
ripple-carry addition circuit by 
Takahashi \cite{cuccaro04,takahashi,takahashi09,parent17} that minimizes both space and gate complexities. 
As it happens,  there are some further improvements and simplifications
 that arise naturally for our particular exponentiation context, which we also exploit. 
All of this is described in detail in Secs.~\ref{subsec:mult} and~\ref{sec:analysis}, 
wherein we also derive fairly accurate resource estimates for both qubit and  Toffoli counts,
respectively.

One disadvantage of  H\"aner multiplication  is that it does not
overwrite the multiplier input---the way, e.g., that QFT multiplication would \cite{draper00,florio04,florio04b}. 
Consequently, each successive multiplication requires additional ancilla bits, unlike what is
presumed in Fig.~\ref{fig2}.  Space needs are accordingly greater in this implementation
than what is described in Sec.~\ref{subsec:ref1}---becoming essentially $m n +d$ qubits rather than 
$2 n+d$ (without ancilla). For the NISQ applications of interest here, $m$ is still quite small, and so the
increase is generally not  too onerous.  It is more of a concern for the Gaussian 
evaluations, for which $m$ can in principle get twice as large as the corresponding 
exponential $d$ value. 

Of course, it would be possible to employ QFT-based multiplication  in  our
exponentiation algorithm---which would require $4n +d$ qubits, with ancilla included. 
On the other hand, the QFT approach is not Toffoli-based, and would therefore not lend
itself to direct comparison with  H\"aner, vis-\`a-vis 
gate complexity. In order to estimate a Toffoli count 
for QFT multiplication, one would have to presume some specific  implementation for the
Toffoli gate itself (e.g., in terms of  \textsc{T} gates), which is not ideal \cite{parent17}.
In any event, Toffoli counts have become a standard gate complexity metric in quantum 
computing. 

For these reasons, overwriting QFT-based multiplications are not considered further here.
Instead, for cases where the increased space complexity of the non-overwriting multipliers
 might pose a problem, we  address this situation through the use of a simple alternative 
 algorithm, describe in Sec.~\ref{subsec:alternative}, that trades increased gate complexity for 
reduced space complexity---essentially by uncomputing intermediate results.
In principle, there are any number of ``reversible pebbling strategies'' \cite{parent17,haner18,gidney19,bennett89} 
that might also be applied towards this purpose.   The particular approach adopted here, though, 
is  very simple, and appears to be quite effective. 


\subsection{\label{subsec:mult} Non-overwriting controlled quantum mutiplication}

\begin{figure*}[htb]
 \centering
 \includegraphics[height=16cm]{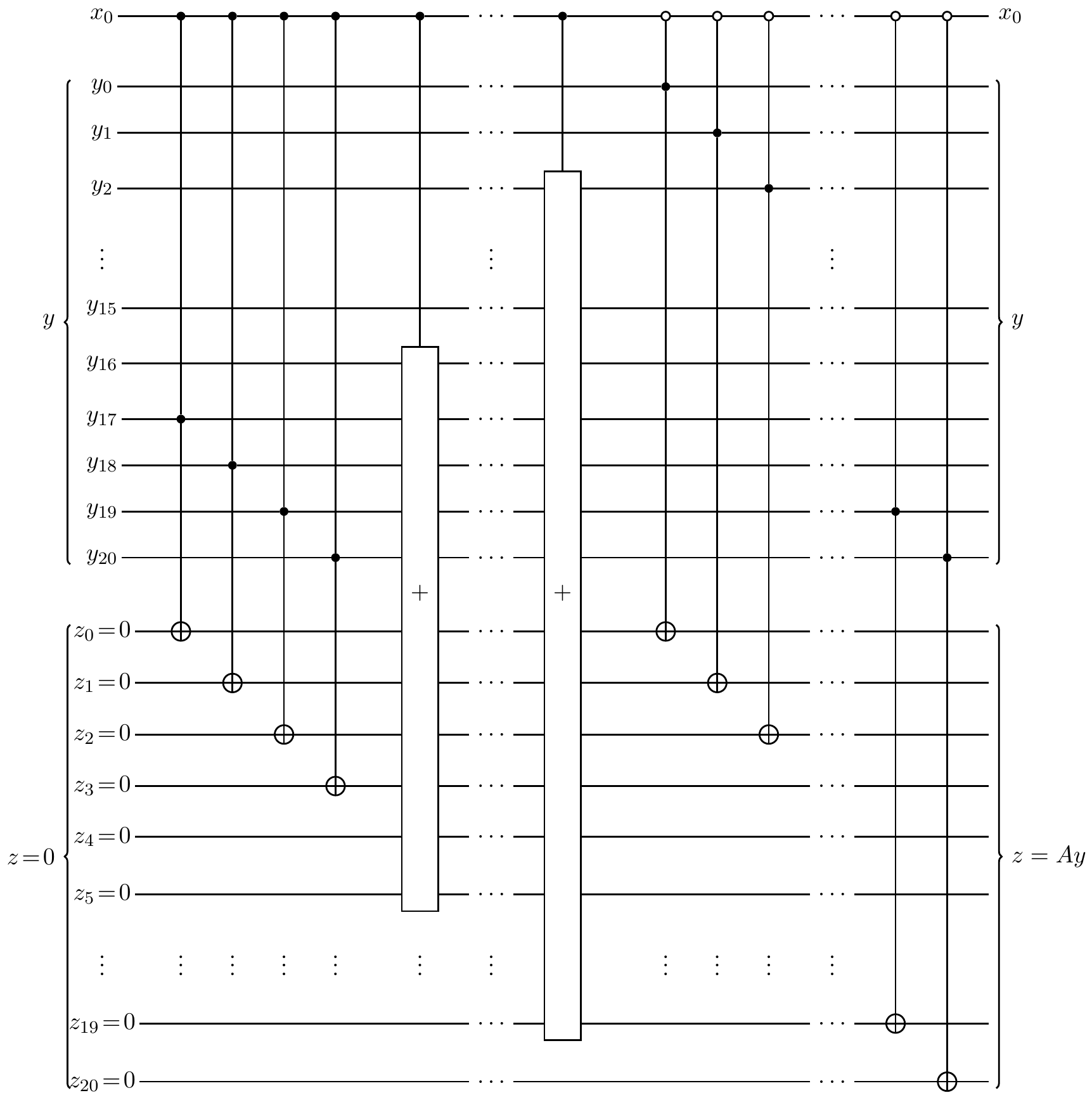}
 \vskip-5pt
 \caption{Detailed Implementation of controlled multiplication on a quantum
computer, $\times_0$, for specific parameter values $d=7$ and $n=21$, 
fixed multiplier, $A_0 = A = 0.389$, and arbitrary superposed multiplicand, $y$. 
The binary representation of $A_0$ is
 $A_0=.011000111001010110000$; each bit with value 1 is hard-wired 
 into the quantum circuit as a distinct controlled addition, $+$.  
 All operations are controlled by the single domain qubit, $x_0$.}
 \label{fig6}
\end{figure*} 

As discussed, non-overwriting controlled-addition multiplication subroutines 
have three registers. The first is an input register for the multiplier; the second
is another input register for the multiplicand; the third is the output or ``accumulator''
register. The accumulator register is initialized to zero, and therefore serves as
an ancilla register,  but 
comes to store the product of the multiplier and  multiplicand at the end of the calculation. 

For integer multiplication, the accumulator register requires $2n$ qubits, assuming
that both input registers are $n$ qubits each. The first register (multiplier) provides the
the control qubits for a cascade of $n$ controlled
additions. The second register (multiplicand) serves as
the first input for each controlled addition.  The second input for each controlled
addition is a successively bit-shifted subset of $n+1$ qubits from the accumulator register.
Note that \emph{overwriting} controlled additions are used, so that for each 
controlled addition, the second register output is the sum of the two inputs.  

In the case of our exponentiation algorithm, we propose a version of the above
basic scheme that is modified in two very important ways. First, the $i$'th multiplication
is  \emph{controlled}, via the domain qubit $x_i$ (Fig.~\ref{fig2}).  Second, the  circuit
exploits the fact that every  multiplier has a \emph{fixed} (unsuperposed) value---i.e. the 
constant, $A_i$. Adding an overall control to a quantum circuit tends  to complicate that circuit---turning 
\textsc{CNOT} gates into \textsc{CCNOT} gates, etc.  On the other hand, the fixed
multiplier enables substantial simplifications---of the type used  in Shor's algorithm for 
factoring integers \cite{parent17,shor94}, for instance.

Specifically, we no longer treat the multiplier $A_i$ as an input register---for there is 
no longer a need to use the $A_i$ qubits as control bits for the  additions. 
Instead, the binary expansion of $A_i$ from \eq{ybin}  is used to \emph{hard-wire}
what would be a set of \emph{uncontrolled} additions, directly into the quantum circuit---but
only for those binary expansion coefficients equal to 1. In addition to reducing the set of inputs 
by one entire $n$-bit register, this modification also reduces the number of additions that must be 
performed  by a factor of two---since on average, only half of the expansion coefficients have 
the value 1. 

In addition to the above advantages, fixed-multiplier  multiplication   reduces circuit complexity
by replacing controlled with uncontrolled additions---effectively converting \textsc{CCNOT}
gates  to \textsc{CNOT} gates.  Of course, when the $x_i$ qubit control is
thrown back in, to create the  requisite \emph{controlled} multiplication subroutine, we find that
the additions become controlled after all---but by $x_i$, rather than  $A_i$.  In effect, the
control bit for the multiplication is simply passed down to the individual controlled additions 
 which comprise it.

The above can all be seen in Fig.~\ref{fig6}, our  detailed quantum circuit for controlled
multiplication, as implemented for the first multiplication  in Fig.~\ref{fig2}, denoted `$\times_0$'
(i.e.,  multiplication by $A_0$,  controlled by the  $x_0$  qubit). Note that the individual 
multiplication circuits, $\times_i$, differ from each other, due to  
the different $A_i$  binary expansions. 
Once again, our  canonical NISQ  parameter values are presumed, i.e., 
$d=7$, $n=21$, and $A=0.389$. 

From the figure, another important difference from the basic scheme may be observed:
the accumulator register, $z$, has only $n$ rather than $2n$ qubits.
This is because fixed-point rather than integer arithmetic is being used---as a consequence of
which, it is not necessary to store what would otherwise be the $n$ least significant bits of
the product.  This situation provides yet another benefit, which is that each controlled addition 
becomes ``truncated'' to an $s$-bit operation---with $s$ increasing with each successive
controlled addition across the range,  $2 \le s \le (n-1)$.  

Note that the smallest possible addition corresponds to $s=2$ rather than $s=1$.
This is because the first controlled addition can be replaced with a cascade of Toffoli 
gates---or controlled bit-copy operations---which is a much more efficient implementation. 
This substitution works because the accumulator
register $z$ is set to zero initially. The very first controlled addition thus always (conditionally) adds the 
multiplicand register $y$  to  zero.  

For the example in the figure, the first four  binary expansion coefficients for $A$ (from right to left) 
are all zero; these bits are simply ignored. The first coefficient equal to one is the $j=4$ or fifth bit. As 
indicated in Fig.~\ref{fig6}, this causes the last four bits of the multiplicand register $y$ 
to be (conditionally) copied into the  first four bits of the accumulator register---in  what would otherwise
be an $s=4$ controlled addition.   
The $j=5$  bit is also equal to one, leading to the first \emph{bona fide} controlled addition
in Fig.~\ref{fig6}, with $s=5$.  This pattern continues until the the next-to-last, or $j=19$ bit is reached, 
which is the last bit equal to one.  This leads to the final controlled addition, with $s=19$.  

Although the last ($j=20$) bit is zero, even if it were equal to one, 
the corresponding controlled addition gate would extend only up to $y_1$. Thus, the top wire, $y_0$,
or least-significant bit of the multiplicand, is never used. This reflects the fact that both numbers
being multiplied have values less than one, and that $n$ fixed  bits of precision are maintained
throughout the calculation. Note also that, as a result, there are never any overflow errors.

The final part of the controlled multiplication circuit is a cascade of $n$ controlled bit-copy operations
(i.e., modified Toffoli gates), which conditionally set the final output of the accumulator register equal to $y$, 
when  $x_0=0$ (hence the open circles).  Otherwise, this register would remain zero. Thus, the ``do nothing'' 
instruction in Sec.~\ref{subsec:basicquantum} does not literally mean ``do nothing'' when non-overwriting
multiplications are used, as it is still necessary to copy the multiplicand input register to the accumulator output
register.

\subsection{\label{subsec:alternative} Quantum algorithm for exponentiation: space saving alternative}

Now that the specific, controlled quantum multiplication algorithm of Sec.~\ref{subsec:mult} has been
identified, we can  determine a precise estimate of space requirements  for 
our overall exponentiation circuit. (Gate complexity will be discussed in Sec.~\ref{subsec:ours}).  
As noted, each of the $m$ multiplications requires a clean $n$-qubit ancilla  bundle as input for its
accumulator register, together with the (accumulator) output from the most recent multiplication 
 as input for its multiplicand register.  Thus, for $m$ successive multiplications, 
$m+1$ separate registers would be required in all.  

\begin{figure*}[htb]
 \centering
 \includegraphics[height=17cm]{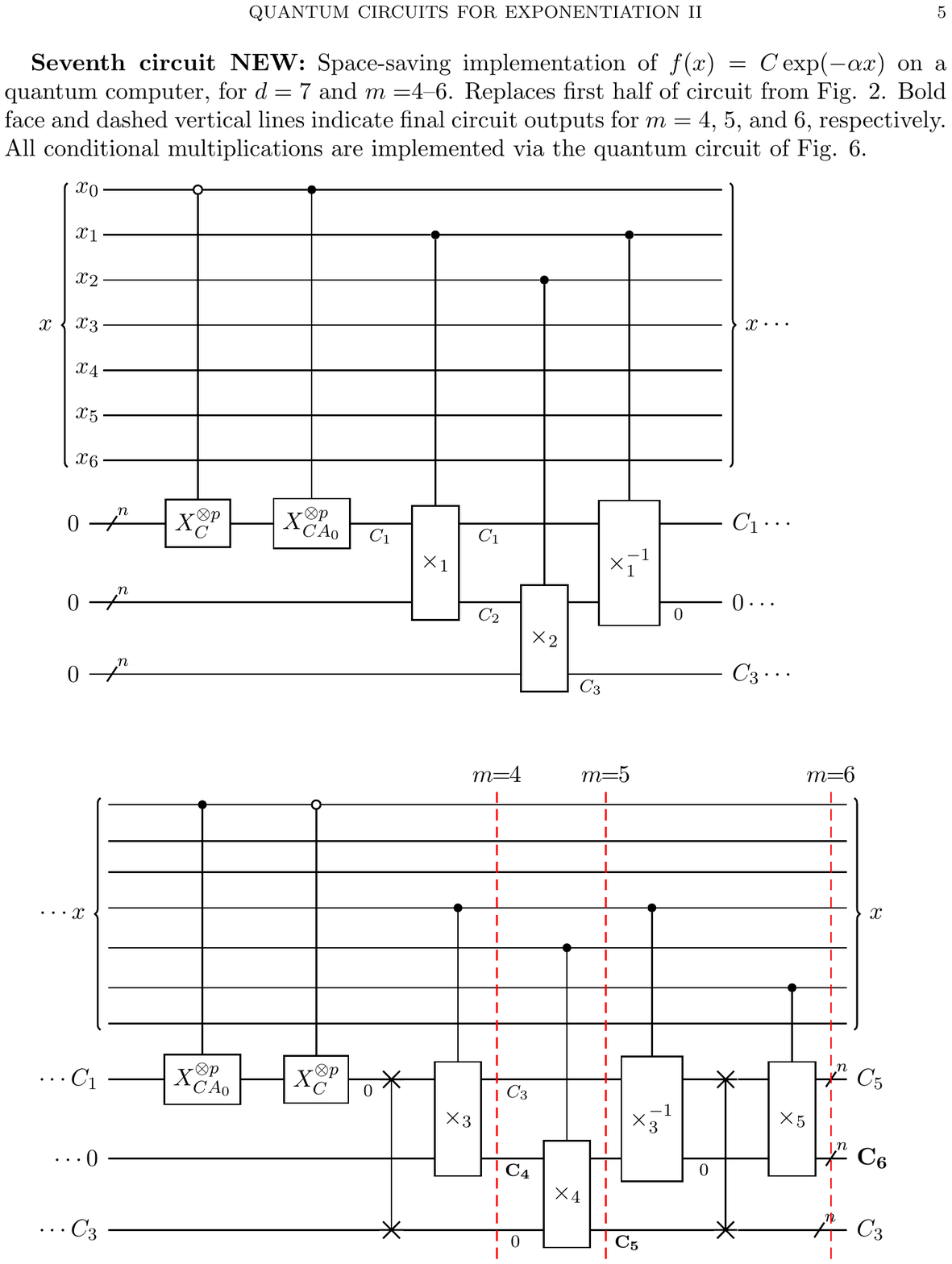}
 \vskip-5pt
 \caption{Quantum circuit used to implement space-saving alternative quantum algorithm for exponentiation, 
 $f(x') = \exp(-\alpha x')$, for specific parameter values,  $d=7$ and $m=$4--6, corresponding to $r=3$.
 Replaces first half of earlier refined circuit, i.e. Fig. 2.  Bold face and dashed vertical lines indicate 
final circuit outputs for $m=4$, 5, and 6, respectively.  All controlled multiplications are 
 implemented via the overwriting quantum circuit of Fig. 6.}
 \label{fig7}
\end{figure*}

However, we can realize significant savings---i.e., one entire register of space, and one
entire controlled multiplication subroutine---by exploiting the fact that the first multiplicand (i.e., $C$) 
is a fixed constant.  The first controlled  multiplication, $\times_0$, is therefore a controlled multiplication of the constant
$C$ by the constant $A_0$.  Since both constants are fixed, the  controlled multiplication can be 
much more efficiently realized using two controlled transformation gates acting on a single 
register (i.e., the first two gates shown in Fig.~\ref{fig7}) rather than the controlled multiplication
circuit of Fig.~\ref{fig6}. Note that this controlled $\times_0$ implementation uses only \textsc{CNOT} 
gates; thus the Toffoli count is zero. 

Since Takahashi addition does not use additional ancilla qubits \cite{takahashi,takahashi09},
the total number of qubits required to implement the $m$ multiplications is just $mn$. In addition to this, 
we have the $d$ qubits needed to store the domain register, $x$, that is used to supply the control qubits. 
The current qubit count  is thus $mn +d$.

However, if   $m<d$, then the  second half of the refined exponentiation circuit (i.e., Fig.~\ref{fig4}) must 
also be executed, which introduces  some additional space overhead.  To begin with, our current reliance 
on non-overwriting  multiplications implies that we can no longer generate the requisite zero
ancilla register (i.e., the  next-to-last register in the figure) without significant (un)computation.  
To avoid this, we instead  add a clean new register---at the additional cost of $n$ new qubits. 
In addition to this, there are the $(d-m-1)$   ancilla bits used by the \textsc{OR} 
gates, as discussed in Sec.~\ref{subsec:second}). Altogether then, the total qubit count becomes:
\eb
	q = \left \{ \begin{array}{ll}
	                 dn+d  & \mbox{for $m=d$} \\ 
	                 (m+1)n + 2d -m-1  &  \mbox{for $m<d$}
	                 \end{array}  \right .
        \label{qcost} 
\ee

To reduce qubit counts in cases where \eq{qcost} renders a NISQ calculation unfeasible, we have developed
a ``space-saving'' alternative algorithm. The general idea
is to uncompute some of the intermediate quantities, in order to restore some of the ancilla registers to their 
initial clean state, so that they can then be reused for subsequent computations. Of course, this requires 
additional overhead---i.e., in our case, additional controlled multiplications. 

More specifically, our  space-saving algorithm reduces  space requirements from $O(mn)$ down to 
$O(m^{1/2} n)$---a very marked reduction, especially if $m$ is reasonably large. The added cost
in terms of gate complexity, on the other hand, is \emph{never more than double} that of our original
algorithm described above. Thus, $m < m_{{\rm ss}}< 2 m$, with $m_{{\rm ss}}$ the multiplication count for the
space-saving approach. 

For values of $m$ that lie  in the range
\eb
	r(r-1)/2 < m \le r(r+1)/2   \label{rcount}
\ee
(where $r>2$ is an integer), the space-saving method requires a total of $r$ $n$-qubit registers to perform
all multiplications. Note that the $r>2$ restriction implies that the method is only applicable for 
$m>3$. However, the $m\le3$ case presents minimal space requirements, and so the space-saving 
approach is less likely to be needed. In any event, for all numerical examples considered in Sec.~\ref{subsec:Haner}, 
(including the worst Gaussian  cases), $ 3 \le r \le 5$. 

The total qubit count for the space-saving alternative algorithm can be shown to be as follows:
\eb
	q_{{\rm ss}} = \left \{ \begin{array}{ll}
	                r n + d & \mbox{for $m=d$} \\ 
	                (r+1)n + 2d -m-1 & \mbox{for $m=r(r+1)/2<d$} \\
	                rn + 2d -m-1 & \mbox{otherwise}
	                \end{array} \right .  \label{qsscost} 
\ee
Note that unlike our original  non-space-saving or ``gate-saving''  algorithm, 
a zero $n$-qubit ancilla register can always be made 
available for the final controlled $\textsc{SWAP}^{\otimes n}$ operation of Fig.~\ref{fig4}---\emph{except}
when  $m=r(r+1)/2$,  which thus has an additional qubit cost.  (See  technical note at the
end of this subsection).

The space-saving algorithm itself proceeds as follows.
First, apply the first $r$ multiplications, exactly as for the earlier gate-saving
algorithm.  This leaves the $r$ registers in the states, $C_1$, $C_2$, \ldots, $C_r$.
Then, uncompute all but the most recent multiplication (i.e., the one that provided $C_r$). The first $(r-1)$ registers
are thereby restored to zero, but the final register remains in the $C_r$ state. It is then possible to perform 
$(r-1)$ additional multiplications, before once again running out of registers.  All but the last of these is
 then uncomputed, allowing $(r-2)$ more multiplications to be performed, and so on.   
 
 The space-saving quantum circuit used for $d=7$ and $m=$4--6 is presented in Fig.~\ref{fig7}, 
 corresponding to $r=3$ registers. 
 For the first wave, there are  three clean registers, allowing for three successive multiplications, 
 $\times_0$, $\times_1$, and $\times_2$ (provided $\times_0$ is implemented as discussed above).
 This is followed by two uncompute multiplications for the first two multiplications, denoted $\times_1^{-1}$ and
 $\times_0^{-1}$ (the latter, again with the new implementation). 
 In the second wave, we apply $\times_3$ and $\times_4$, generating $C_4$ and $C_5$,
 respectively.  This suffices for $m=4$ and $m=5$, respectively.  However, if $m=6$, we 
 must undergo a third and final wave, as indicated in the figure. 

As is clear from the above description, and from Fig.~\ref{fig7}, the number of uncompute
multiplications, $m_{{\rm un}}$ ,  is always less than $ m$. Thus, $m_{{\rm ss}} = m + m_{{\rm un}} < 2m$,
as claimed.
Precise values can be found as follows. Let $l$ be the largest integer such that
\eb
	l(l+1)/2 \le r(r+1)/2 - m. \label{lcond}
\ee
Then, 
\eb
	m_{{\rm un}} = r(r-1)/2 - l(l+1)/2.
\ee

Table~\ref{mtab} indicates specific values for all $m\le36$. Note that the $m_{{\rm ss}}$ multiplication
count includes 
both $\times_0$ and $\times_0^{-1}$; thus, the total \emph{actual} number of controlled multiplication 
subroutines that must be executed is $(m_{{\rm ss}}-2)$, as indicated in the final column. From the table,
also, it may be observed that greater space savings are usually associated with increased  
multiplication counts, and vice-versa.

\emph{Technical note:} 
For $m < d$ space-saving calculations, a zero ancilla register is automatically available 
at the end of the Fig.~\ref{fig7} circuit (to be used in the subsequent Fig.~\ref{fig4} circuit), 
whenever \eq{lcond} is a true inequality.  When \eq{lcond} is an \emph{equality}, then the $l=0$ case
requires the addition of a new zero ancilla register (as discussed), but  for $l>0$, a zero register can be easily 
created from an existing non-zero register. This is done by applying the single uncompute multiplication,
$\times_{m-2}^{-1}$.  As an example, the case $m=5$ corresponds to $l=1$ and $r=3$, 
thus satisfying \eq{lcond} as an equality, with both sides equal to one.  The necessary 
uncompute multiplication can be seen in Fig.~\ref{fig7}, just to the right of the vertical dashed line 
marked `$m=5$'.  Note that the Toffoli count associated with such $i \approx m$  multiplications is  
greatly reduced in comparison with the other multiplications, as will be discussed in Sec.~\ref{subsec:ours}.  
Moreover, this event is fairly rarely realized in practice, including the examples given in the present work. 
Nevertheless,  the small additional cost required in such cases is included in the Toffoli count formulas presented
in Sec.~\ref{subsec:ours}.

\begin{table}
\centering
\caption{Number of $n$-qubit registers $r$, uncompute multiplications $m_{{\rm un}}$ (including
$\times_0^{-1}$), and total actual multiplications $(m_{{\rm ss}}-2)$, as a function of number of 
compute multiplications $m$ (including $\times_0$), for the
space saving alternative quantum exponentiation  algorithm of Sec.~\ref{subsec:alternative} .}
\begin{tabular}{ p{1cm} p{1.25cm} p{1.25cm} p{1.3cm} p{1.25cm} p{1.25cm}}
\hline
\hline
$m$ & $r$ & $m_{{\rm un}}$ & $(m_{{\rm ss}}-2)$ \\
\hline
 4 & 3 & 2 & 4 \\
 5 & 3 & 2 & 5 \\
 6 & 3 & 3 & 7 \\
 7 & 4 & 3 & 8 \\
 8 & 4 & 5 & 11 \\
 9 & 4 & 5 & 12 \\
 10 & 4 & 6 & 14 \\
 11 & 5 & 7 & 16 \\
 12 & 5 & 7 & 17 \\
 13 & 5 & 9 & 20 \\
 14 & 5 & 9 & 21 \\
 15 & 5 & 10 & 23 \\
 16 & 6 & 12 & 26 \\
 17 & 6 & 12 & 27 \\
 18 & 6 & 12 & 28 \\
 19 & 6 & 14 & 31 \\
 20 & 6 & 14 & 32 \\
 21 & 6 & 15 & 34 \\
 22 & 7 & 15 & 35 \\
 23 & 7 & 18 & 39 \\
 24 & 7 & 18 & 40 \\
 25 & 7 & 18 & 41 \\
 26 & 7 & 20 & 44 \\
 27 & 7 & 20 & 45 \\
 28 & 7 & 21 & 47 \\
 29 & 8 & 22 & 49 \\
 30 & 8 & 22 & 50 \\
 31 & 8 & 25 & 54 \\
 32 & 8 & 25 & 55 \\
 33 & 8 & 25 & 56 \\
 34 & 8 & 27 & 59 \\
 35 & 8 & 27 & 60 \\
 36 & 8 & 28 & 62 \\
\hline
\end{tabular}
\label{mtab}\\
\end{table}

\section{\label{sec:analysis} Analysis: Toffoli counts}

\subsection{\label{subsec:ours} Present methods}

To a rough approximation, the total Toffoli count for the proposed exponentiation algorithm
[or for the space-saving alternative]  is simply  $(m-1)$ [or $(m_{{\rm ss}}-2)$] 
times the Toffoli count needed to execute  a single controlled multiplication subroutine.  
Before working out the latter, however, we first describe another reduction of effort that in
practical terms, converts the total cost to that of only $(m-8/3)$ [or $(m_{{\rm ss}}-16/3)$]
controlled multiplications. This additional savings is fully realized 
whenever $m<d$, which in practice occurs much of the time, if the domain interval is
realistically large.

The rationale is as follows. When $m <d$, the $A_i$ values \ span the 
entire range from 1  down to $2^{-n}$. This implies that for  $i \approx m$, 
the corresponding $A_i$  have many leading zeroes. Consequently, these later 
multiplications can be performed using  fewer than $n$ binary digits, leading to 
significant computational savings.

Note that  the \emph{worst-case} scenario vis-\`a-vis the aforementioned savings---i.e., 
that for which the  $A_{i\approx m}$ have the \emph{fewest} leading zeros---corresponds to 
$A_m \rightarrow 2^{-n}$ from below.
Now, in general, the approximate number of leading zeros for the binary expansion
of $0 < y <1$ is given by  $-\log_2 y$.  Thus, for $y=A_m$, we find $\sim \!\!n$ leading
zeros, as expected. More generally, \eq{Ai} in the worst-case scenario leads to
\eb
	\mbox{leading zeros}(A_i)  \approx   n\,  2^{-(m-i)} = n \, 2^{-k},
\ee
where $k = (m-i)$.  
The number of binary digits needed for the $A_{m-k}$ controlled multiplication is thus 
$n_k = n - n \, 2^{-k}$.  

Going forward, we shall for simplicity presume the asymptotic limit, $n \ra \infty$.
In this limit, the Toffoli count per multiplication scales as $O(n_k^2)$. The Toffoli
\emph{savings} (i.e., reduction in the Toffoli count relative to multiplication with 
$n$ digits) is therefore
\eb
	s_k  \propto (n^2 - n_k^2) =  n^2 (2 \, 2^{-k} - 2^{-2k})  \label{savings}
\ee
 Summing \eq{savings} from $k=1$ to $\infty$ then yields a total savings of $5/3$
 multiplications. 
 
 In practice---i.e., for finite $n$---the series is truncated, and so the actual 
 savings is less than $5/3$ multiplications.  In the worst case (of the worst case), 
 only the $s_1$ term contributes to the sum, resulting in a lower bound of $3/4$
multiplications.  On the other hand, a small increase in $A_m$, such that the 
new value is slightly greater than $2^{-n}$, will increment the value of $m$---thus, 
effectively increasing the savings by one whole additional multiplication. On
balance, we therefore take our $5/3$
 ``best case of the worst case'' value as a reasonable middle-ground estimate.   
 
Next, we move on to a calculation of the Toffoli cost of each controlled multiplication.
As discussed in Sec.~\ref{subsec:mult}, these are implemented using a sequence
of controlled additions, with from $s=2$ qubits up to $s=(n-1)$ qubits. Note that 
the Toffoli cost of the highly efficient overwriting, controlled, ripple-carry addition 
circuit of Takahashi \cite{takahashi,takahashi09,haner18} with $s$ 
qubits is $3s+3$. Thus, if \emph{every} $2\le s \le (n-1)$ required a controlled 
addition, the total  contribution to the Toffoli cost of multiplication would be 
$(3/2)n^2 + (3/2) n -9$. However, since only \emph{half} of these multiplications
are realized on average, in practice, the actual cost per multiplication is half of this. 
The total contribution to the cost of the exponentiation circuit is then this value, multiplied 
by the \emph{effective} number of multiplications, i.e. $(m-8/3)$.

Now, in addition, each controlled multiplication in the Fig.~\ref{fig6} circuit also
begins and ends with a cascade of additional Toffoli gates.   The initial cascade
can be easily shown to consist of two Toffoli gates, on average. The final cascade,
is always $n$ Toffoli gates, even when fewer than $n$ qubits are needed to execute
the main part of the multiplication circuit (i.e., the controlled additions). Note that
both Toffoli cascades are required in every \emph{actual} multiplication. The 
total contribution to the Toffoli cost of the exponentiation circuit is thus $(m-1)(n+2)$.

Finally, there are the additional costs associated with the second half of the (refined)
exponentiation circuit, as presented in Fig.~\ref{fig4}, presuming $m <d$.  Since
each Fredkin gate can be implemented using a single Toffoli gate, the Toffoli cost of the
final $\textsc{SWAP}^{\otimes n}$ operation is $n$. Likewise, each \textsc{OR} gate
requires one Toffoli gate, for a total Toffoli count of $(d-m-1)$. Altogether, we 
wind up with the following expression for the total Toffoli cost for the entire 
gate-saving exponentiation circuit: 
\eb
	T = \left \{ \begin{array}{ll}
	                 \of{{3d \over 4}\!-\!2}  n^2  + \of{{7d \over 4}\!-\!3}  n  - {5 \over 2} d + 10  & \mbox{for $m=d$} \\ [2mm]
	                \of{{3m \over 4}\!-\!2}  n^2  + \of{{7m \over 4}\!-\!2}  n  - {7 \over 2} m + d +9 &  \mbox{for $m<d$}
	                 \end{array}  \right .
        \label{Tcost} 
\ee

Things are a bit more complicated in the space-saving algorithm case. In particular, there are
three cases instead of two. In addition to $m=d$, there are two different $m<d$ cases, i.e. one
corresponding \eq{lcond} being an equality, and one to the inequality case,  as discussed in the 
Technical note at the end of Sec.~\ref{subsec:alternative}.  Note also that the uncompute
multiplications that are \emph{not} included in $m_{{ \rm un}}$ (as compared to $m$) are in
fact the $i \approx m$ multiplications, that do not cost as much.  Consequently, the  effective
number of uncompute multiplications is reduced relative to the actual number, by an amount
\emph{less than} 5/3 multiplications. A more accurate estimate of the uncompute savings is given by
\eb
	S_{\Delta m} =  \sum_{k=1+\Delta m}^{\infty} s_k\! = \!n^2\of{ 2\, 2^{-\Delta m} - 2^{-2 \Delta m}/3},
\ee 
where $\Delta m = (m - m_{{\rm un}})$. 

Taking all of the above into account, we obtain the following expression for the Toffoli count of 
the space-saving exponentiation algorithm:
\eb
	T_{{\rm ss}} = \left \{ \begin{array}{ll}
		\of{m_{{\rm ss}}\! -\! {11 \over 3}  -S_{\Delta m}}\of{{3 \over 4}n^2 \!+ \!{3 \over 4} n \!-\! {9 \over 2}} 
					+ (m_{{\rm ss}} \!-\!2)(n\!+\!2)  \\ 
				\hskip 6.0cm \mbox{for $m=d$}  \\ [2mm]
		\mexp{\mbox{above}} + n+d-m-1 \hskip 2.2cm  \mbox{for $m<d $} \\ [4mm]
	         \mexp{\mbox{above}} + {9 \over 16}\of{{3 \over 4}n^2 \!+ \!{3 \over 4} n \!-\! {9 \over 2}} + (n\!+\!2) \\ [0.5mm]
		 \hskip 0.1cm \mbox{for $m\!<\!d$ and $l\!>\!0$ and $l(l\!+\!1)/2 = r(r\!+\!1)/2\! -\! m$}  \\ 	         
	                 \end{array}  \right .
        \label{Tsscost} 
\ee
Note that in the final case above, the (worst-case) cost of the additional $\times_{m-2}^{-1}$
uncompute multiplication is obtained from $s_2$ in \eq{savings} to be $9/16$ that of a regular
multiplication---at least insofar as the controlled addition contribution is concerned.

\subsection{\label{subsec:Haner} Explicit numerical comparison with H\"aner approach}

In the approach by H\"aner et al. \cite{haner18}, arbitrary functions are evaluated via a decomposition
of the $x'$ domain into $M$ non-intersecting subdomain intervals, as discussed in Sec. ~\ref{sec:Intro}.
A given  function is then approximated
using a separate $d$'th order polynomial in each subdomain.  Both the polynomial coefficients,
and the subdomain intervals themselves, are optimized for a given target accuracy, using the
Remez algorithm \cite{remez34}.  

Once the optimized parameters have been determined
for a given function $f(x')$, domain interval $x'_{{\rm min}} \le x' <x'_{{\rm max}}$, and (H\"aner) $d$ value, 
the quantum algorithm is then implemented as follows. 
First, polynomials are evaluated using a sequence of $d$ multiplication-accumulation
(addition) operations.  On a quantum platform,  these can be performed in parallel, across 
all $M$ subdomains at once, using conditioned determination of the coefficients for each 
subdomain.  The multiplication count
would thus be $d$, \emph{irrespective} of $M$.  Also, since non-overwriting multiplications are used,
the qubit count is $O(nd)$. 

Note that generally speaking, lower $d$ corresponds to greater $M$, and
vice-versa.   Thus, were the above multiplication-accumulation operations the only significant computational cost, 
one would simply choose a very small value such as $d=1$.  However, there is additional 
space and gate complexity overhead associated with managing and assigning the $M$ sets of polynomial
coefficients.  These costs do increase with $M$ (although in a manner that is  naturally
measured in Toffoli gates rather than multiplications).  There is thus a competition between $M$ and $d$,
with minimal Toffoli counts resulting when $d=4$ or 5---at least for the numerical examples from 
\Reff{haner18} that are considered here. 

The minimal-Toffoli choice of $d$  can thus be thought of as  a ``gate-saving'' H\"aner 
implementation.
Note that a rudimentary ``space-saving'' alternative may also be  obtained, simply by reducing the  value of $d$. That
said, \Reff{haner18} also discusses the use of much more sophisticated pebbling strategies. 
 However, such strategies are not actually implemented for
the numerical results presented in \Reff{haner18} that are used for comparison with the present
results.

Instead,  H\"aner and coworkers perform calculations for different functions,  and for different 
target accuracies, across a range of different $d$  values---providing total  qubit and Toffoli 
counts for each. In particular, they consider both the exponential and Gaussian functions, 
with $x'_{{\rm min}}=0$ and $\alpha=1$. It  thus becomes mostly possible to provide a direct 
comparison between the H\"aner approach and our
methods, with respect to these metrics. Such a comparison is provided in Table~\ref{restab}. 

One slight difficulty arises from the fact that \emph{no $x'_{{\rm max}}$ value is provided in \Reff{haner18}};
moreover, the authors have not been available for clarification.  
We thus present results for our methods using two very different $x'_{{\rm max}}$ values---i.e., 10
and 100.  Although for most purposes, even the former interval is wide enough to capture the main  
function features, the latter interval is actually more realistic for certain applications such as QCC
(as discussed in greater detail in Sec.~\ref{sec:conclusions}).  In any case, it should be mentioned
that in the large $x'_{{\rm max}}$ limit, our methods become less expensive, whereas the H\"aner approach
 becomes more expensive---owing to the increased $M$ values needed to achieve a given
level of accuracy. 

The issue of target range accuracy  merits further discussion. In H\"aner, calculations were performed
to an accuracy of $10^{-7}$, and also $10^{-9}$.  The  corresponding number of bits needed to resolve the
range to these thresholds are 23.25 and 29.89, respectively.  Note that these values are quite close to the $n=21$
and $n=32$ values considered in our examples thus far---in one case a bit high, in the other a bit low. Of 
course, a few extra bits might also be needed to compensate for round-off error in the fixed-point arithmetic. 
These are relatively small effects however; in particular, they are likely no larger than those associated with the unknown 
H\"aner $x_{{\rm max}}$ value. 

For our purposes, therefore, we take these $n$ values as reasonable
estimates, at least for the exponential function evaluations. For the Gaussian case, we do go ahead and use
$n=24$ and $n=30$, as the closest  integers larger than the threshold values listed above.  As for the
corresponding $d$ values, roughly speaking, these  would be double those from the exponential 
calculation---except that we exploit symmetry of the domain to reduce these by one bit each. 
Thus, $d=13 = 2 \times7 -1 $ and $d= 15=2 \times 8 -1 $, respectively, for $n=24$ and $n=30$.

In Table~\ref{restab}, we present Toffoli and qubit counts for both function evaluations (i.e., exponential and Gaussian),  
for both target accuracy thresholds (i.e. $10^{-7}$ and $10^{-9}$), for both of our methods (i.e. gate-saving
and space-saving), and for both domain intervals (i.e.  $0 \le x' < 10$ and  $0 \le x' < 100$). For each 
function,  the minimal Toffoli count is given in bold face.  For each function and accuracy
threshold, we also present results for the full set  of H\"aner calculations, as obtained 
from \Reff{haner18}.  Here too, the minimal Toffoli count is highlighted in bold face. 

In all cases, our method requires \emph{far} fewer Toffoli gates than the H\"aner approach.
 In comparing  minimal-Toffoli calculations for the exponential function, the Toffoli count is reduced from
15,690 using H\"aner, down to just 912 using our approach. For Gaussian function evaluation, the 
Toffoli count comparison
is even more stark---i.e., 19,090 vs. just 704.   Generally speaking, our methods also require fewer qubits
than H\"aner. This is especially true for the space-saving alternative, which in one instance requires
as few as 71 qubits (and 1409 Toffoli gates)---a NISQ calculation, certainly, by any standard.

\begin{table*}
\centering
\caption{Toffoli  and qubit counts for exponential [$\exp (-x')$] and Gaussian [$\exp (-{x'}^2)$] function 
evaluations, using three different methods: gate-saving (ours); space-saving (ours); H\"aner. 
For our methods, two different domain intervals are used: $0 \le x' < 10$ and  $0 \le x' < 100$.
Two different target accuracies are considered: 
$10^{-7}$ (Columns IV--VI) and $10^{-9}$ (Columns VII--IX).
For the former, bold face indicates minimal Toffoli count  from among a given set of calculations,
i.e. ours vs.  H\"aner.}
\begin{tabular}{ccc|rrr|rrr}
\hline
\hline
Function & Method & Domain & \multicolumn{3}{|c}{$10^{-7}$ Accuracy}  & \multicolumn{3}{|c}{$10^{-9}$ Accuracy}  \\
\cline{4-9} 
 & & interval &  $\quad (n,d,m)$  & $\quad$ Toffolis &  $\quad$ qubits   & $\quad(n,d,m)$  & $\quad$  Toffolis & $\quad$ qubits  \\
\hline
$\exp(-x')$   & $\quad$ gate saving $\quad$  & $0\le x' \le 10$  & $(21,7,7)$ & 1620  & 154 & $(32,8,8)$ & 4438 & 264  \\
		   &                          & $\quad 0\le x' \le 100\quad$   & $(21,7,5)$ & {\bf 912}    &   134 & $(32,8,6)$ &  2828 & 233  \\
                    & space saving   & $0\le x' \le 10$                         &$(21,7,7)$ & 2308 &     91 & $(32,8,8)$ & 7531 & 136  \\
		   &                          & $\quad 0\le x' \le 100\quad$   & $(21,7,5)$ & 1409 & 71     & $(32,8,6)$ &  4278 & 105  \\
                    & H\"aner             &                                               &                   & 17304 & 149 &                 & 45012 & 175  \\
                    &                          &                                               &                   & {\bf 15690} & 184 &                 & 28302 & 216  \\
                    &                          &                                               &                   & 16956 & 220 &                 & 25721 & 257  \\
                    &                          &                                               &                   & 18662 & 255 &                 & 26452 & 298  \\
\hline
$\exp(-{x'}^2)$   & $\quad$ gate saving $\quad$  & $0\le x' \le 10$  &$\quad (24,13,12)$ & 4468 & 325 & $\quad(30,15,14)$ & 8300 & 465  \\
		   &                          & $\quad 0\le x' \le 100\quad$       & $(24,13,4)$ & {\bf 704}   & 141  & $(30,15,7)$ &   3232 & 262  \\
                    & space saving   & $0\le x' \le 10$                             &$(24,13,12)$ & 7546 & 133 & $(30,15,14)$ & 14479 & 165  \\
		   &                          & $\quad 0\le x' \le 100\quad$       & $(24,13,4)$ & 962  &  93   & $(30,15,7)$ &  5018 & 142  \\
                    & H\"aner             &                                               &                   & 20504 & 161 &                 & 49032 & 187  \\
                    &                          &                                               &                   & {\bf 19090} & 199 &                 & 32305 & 231  \\
                    &                          &                                               &                   & 21180 & 238 &                 & 30234 & 275  \\
                    &                          &                                               &                   & 23254 & 276 &                 & 31595 & 319  \\
\hline
\end{tabular}
\label{restab}\\
\end{table*}

\section{\label{sec:conclusions}Summary and Conclusions}


After the various refinements and NISQ-oriented details as presented
in the latter 2/3 of this paper, it might be easy to lose sight of the main point,
which is simply this:  \emph{the method  presented here allows the exponential
function to be evaluated on quantum computers for the cost of a few multiplications.}
This basic conclusion will continue to hold true, regardless of the many quantum 
hardware and software innovations that will come on the scene in  ensuing
decades. 

In particular, there is a plethora of multiplication algorithms available, both 
overwriting (e.g. QFT-based) and non-overwriting (e.g. controlled addition), 
and for both integer and fixed-point arithmetic---with new strategies for floating-point arithmetic, 
quantum error correction, etc., an  area of ongoing development.  Given 
this  milieu,  we propose  the implementation-independent ``multiplication count'' 
as the most sensible gate complexity metric, for any  quantum algorithm whose 
dominant cost can be expressed in terms of multiplications. The present algorithms 
are certainly  of this type. 

For our exponentiation strategy, the (controlled) multiplication count $m$ will 
indeed be rather small in practice---at least for the applications envisioned.
 To begin with, in a great many  simulation contexts,
the domain resolution as expressed in total qubits $d$, is far less than the range
resolution $n$---with $m \le d$.  For QCC, for instance, the $d=7$ and $d=8$ values 
considered throughout this work are likely to suffice in practice \cite{mypccp}.
Conversely, we also consider the asymptotically large $n=d$ limit, in which it can 
be shown [in \eq{meq}] that $m=O(\log n)$ for fixed $A$. In this limit, Karatsuba
multiplication provides better asymptotic scaling.  Using the Gidney implementation, 
the Toffoli and qubit counts for exponentiation scale as  $O(n^{\log_2 3} m)$ and 
$O(nm)$, respectively.  

In the latter part of this paper, we  present two specific,  NISQ 
implementations of our general exponentiation strategy, in order  that detailed resource 
estimates can  be assessed, and compared with competing methods. When compared 
with the  method of H\"aner and coworkers, our implementations are found to 
reduce Toffoli counts by an order of magnitude or more.  Qubit
counts are also  (generally) substantially reduced.  Note that  our  two 
 implementations  are complementary,  with one designed to favor gate and the other 
 space resource needs. Together, they may provide the flexibility needed to actually 
implement  exponentiation on NISQ architectures---which 
could serve as the focus of a future project. 

Finally, we assess the present exponentiation algorithms within the context in which
they were originally conceived---i.e., quantum computational chemistry  (QCC). 
The long-awaited ``(QCC) revolution''\cite{poplavskii75,feynman82,lloyd96,abrams97,zalka98,lidar99,abrams99,nielsen,aspuru05,kassal08,whitfield11,brown10,christiansen12,georgescu14,kais,huh15,babbush15,kivlichan17,babbush17,babbush18,babbush18b,babbush19,low19,kivlichan19,izmaylov19,parrish19,altman19,cao19,alexeev19,bauer20,aspuru20}'' 
may be nearly upon us,  although achieving full quantum supremacy will likely  require quantum 
platforms that can accommodate first-quantized methods.  On classical computers, the
Cartesian-component separated (CCS) approach, as developed by the 
author \cite{jerke15,jerke18,jerke19,mypccp,bittner}
offers a highly competitive first-quantized strategy. 

On quantum computers, the question 
appears to boil down to the relative costs of  the exponential function vs. the  inverse 
square root \cite{babbushcomm}.   Toffoli count estimates for the former appear in Table~\ref{restab}.
Note that the larger-domain-interval calculations---i.e.,  those with lower Toffoli counts---are the 
more realistic in this context.  This is because  the QCC CCS
implementation requires multiple exponentiations with different $\alpha$ values to be performed,
across the same grid domain interval---which must accordingly be  large enough to 
accommodate all of them.  Our exponentiation cost of 704 Toffoli gates should thus be
compared to the cost of the inverse-square-root function, which---again, according to the highly optimized 
method of H\"aner and coworkers---is estimated to be 134,302 Toffoli gates. 




\section*{\label{sec:acknw}Acknowledgement}

The author gratefully acknowledges support from a grant from the Robert A. Welch Foundation (D-1523).

\bibliography{references}
\bibliographystyle{rsc}

\end{document}